\documentclass{article}

\usepackage{arxiv}

\usepackage[utf8]{inputenc} 
\usepackage[T1]{fontenc}    
\usepackage{hyperref}       
\usepackage{url}            
\usepackage{booktabs}       
\usepackage{amsfonts}       
\usepackage{nicefrac}       
\usepackage{microtype}      
\usepackage{lipsum}
\usepackage{graphicx}

\title{Exploration of Spanish Olive Oil Quality with a Miniaturized Low-Cost Fluorescence Sensor and Machine Learning Techniques}

\author{
  Francesca Venturini\\
  Institute of Applied Mathematics and Physics\\
  Zurich University of Applied Sciences\\
  Technikumstrasse 9, 8401 Winterthur, Switzerland \\
  \texttt{vent@zhaw.ch} \\
   \And
    Michela Sperti \\
  Polito BIO Med Lab\\
  Department of Mechanical and Aerospace Engineering\\
  Politecnico di Torino, Turin, Italy \\
  \texttt{michela.sperti@polito.it } \\
  \And
  Umberto Michelucci \\
  TOELT LLC\\
  Research and Development \\
  Birchlenstr. 25, 8600 D\"ubendorf, Switzerland  \\
  \texttt{umberto.michelucci@toelt.ai } \\
  \And
  Ivo Herzig\\
  Institute of Applied Mathematics and Physics\\
  Zurich University of Applied Sciences\\
  Technikumstrasse 9, 8401 Winterthur, Switzerland \\
  \And
  Michael Baumgartner\\
  Institute of Applied Mathematics and Physics\\
  Zurich University of Applied Sciences\\
  Technikumstrasse 9, 8401 Winterthur, Switzerland \\
  \And
  Josep Palau Caballero\\
  SCA San Sebastián Puente del Ventorro\\ 
  s/n 18566 Benalua de las Villas, Spain\\
  \And
  Arturo Jimenez\\
  SCA San Sebastián Puente del Ventorro\\ 
  s/n 18566 Benalua de las Villas, Spain\\
  \And
  Marco Agostino Deriu\\
  Polito BIO Med Lab\\
  Department of Mechanical and Aerospace Engineering\\
  Politecnico di Torino, Turin, Italy \\
  \texttt{marco.deriu@polito.it }
}

\begin{document}
\maketitle

\begin{abstract}
Extra virgin olive oil (EVOO) is the highest quality of olive oil and is characterized by highly beneficial nutritional properties. The large increase in both consumption and fraud, for example through adulteration, creates new challenges and an increasing demand for developing new quality assessment methodologies that are easier and cheaper to perform. As of today, the determination of olive oil quality is performed by producers through chemical analysis and organoleptic evaluation. The chemical analysis requires the advanced equipment and chemical knowledge of certified laboratories, and has therefore a limited accessibility. In this work a minimalist, portable and low-cost sensor is presented, which can perform olive oil quality assessment using fluorescence spectroscopy. The potential of the proposed technology is explored by analyzing several olive oils of different quality levels, EVOO, virgin olive oil (VOO), and lampante olive oil (LOO). The spectral data were analyzed using a large number of machine learning methods, including artificial neural networks. The analysis performed in this work demonstrates the possibility of performing  classification of olive oil in the three mentioned classes with an accuracy of 100$\%$. These results confirm that this minimalist low-cost sensor has the potential of substituting expensive and complex chemical analysis.
\end{abstract}


\section{Introduction}

Olive oil is an important commodity in the entire world, and its demand has grown substantially in recent years. The interest in its highest quality grade, extra virgin olive oil (EVOO), is due to its high nutritional value, its richness in bioactive molecules, and its importance to our health due to its content of anti-inflammatory and antioxidant substances \cite{georgiou2017food}. The increased demand has led, however, to an increase in fraudulent activities like adulteration. As a result, edible olive oil quality assessment has become increasingly important. To develop a trusted means of control, the European Economic Community (EEC) has created regulations that define the categorization of olive oils according to several chemical properties, obtainable by accredited laboratories, and organoleptic evaluation, obtainable by accredited panels, to guarantee its quality \cite{regulation1991commission}. 
For the highlighted reasons, the quality control is complex, costly and cannot be carried out easily at any desired moment during the product life cycle.
An inexpensive tool for an accessible analysis will boost consumers' trust in the product and decrease dramatically production costs, reducing at the same time the possibilities for fraudulent activities.

Fluorescence spectroscopy has attracted a lot of interest in recent years as a fast, cost-efficient and at the same time sensitive method to study the properties of vegetable and particularly olive oils \cite{kongbonga2011characterization,sikorska2012analysis}. This is due to the fact that olive oils contain several natural fluorescence molecules like pigments, such as chlorophyll and beta-carotene, phenolic compounds, such as tocopherol, and their oxidation products.
The most frequently used techniques are either the acquisition of excitation emission matrices (EEMs) or the use of synchronous scanning \cite{skoog2017principles}. Both take advantage of the multidimensional characteristic of fluorescence spectroscopy to create a fingerprint to uniquely identify and characterize virgin olive oils  \cite{guzman2015evaluation,meras2018detection}.
Applications of those methods range include discrimination of different quality grades of olive oils \cite{guimet2004cluster,poulli2005classification}, detection of adulteration \cite{sayago2004detection,poulli2007rapid,ali2018validation}, monitoring of the oxidation processes \cite{hernandez2017fast,mishra2018monitoring,baltazar2020development}, shelf-life monitoring \cite{lobo2020monitoring}, and geographical origin authentication \cite{dupuy2005origin,jimenez2019comparative,al2021cultivar}.

The extraction of information of interest from the spectral data can be a difficult task depending on the type of data acquired, which may range from a single spectrum to the more complex EEMs, and on the specificity of the application. Several multivariate analysis techniques and classification methods have been successfully employed, like for example, principal component analysis (PCA), partial least square regression (PLS) and PLS discriminant analysis (PLS-DA), linear discriminant analysis (LDA), K-nearest neighbors (k-NN) and random forest (RF), to mention only the most widely used. More recently the use of artificial neural networks (ANN) has proven to be a useful tool, particularly because it does not require the pre-processing of the data or a dimensionality reduction \cite{Michelucci2017}. Complete overviews of the mentioned statistical and machine learning methods, including ANN, can be found in \cite{sikorska2012analysis,sikorska2014vibrational,zaroual2021application,meenu2019critical,gonzalez2019critical}.

The acquisition of high-quality data, particularly of EEM, and the necessary data post-processing require special instrumentation and knowledge, thus limiting the accessibility of these methods. To the best of the author's knowledge, no portable and low-cost sensors for fluorescence spectroscopy for quality assessment of olive oils are available so far.
This work presents a minimalist sensor for olive oil quality assessment based on fluorescence spectroscopy and shows how it can be used to perform classification without any sample-preparation and without any pre-processing of the acquired data with several machine learning methods.

The main contributions of this paper are three. 
Firstly, a new miniaturized and low-cost fluorescence sensor is presented. The sensor is demonstrated by using it to producing data (fluorescence spectra) that can be used to successfully classify the olive oils samples into three quality classes. 
Secondly, eight different machine learning methods are applied to the data acquired with the sensor, to demonstrate that the data are extremely effective in allowing machine learning models to learn to predict olive oil's quality almost perfectly. A detailed comparison of the models used is discussed. Finally, the performance of ANNs is analyzed in detail. The study of ANNs' performance is an important contribution since ANNs allow the application of explainability techniques to better understand how olive oil quality is linked to its chemical properties. This has the potential of completely superseding the classical chemical analysis.

\section{Materials and Methods}
\label{sec:material_methods}

\subsection{Olive Oil Samples}
\label{sec:OliveOilSamples}

All samples were obtained from the 2019-2020 harvest and provided by the producer Conde de Benalúa, Granada, Spain. In total, 27 olive oil samples were analyzed in this study, divided into 12 EVOO, 8 VOO, and 7 LOO (Table \ref{tab:oils}). 
The quality assessment of all the olive oils was performed by the producer according to the current European regulation for the commercial classification into EVOO, VOO, and LOO categories \cite{regulation1991commission}. The quality is determined by both chemical parameters, such as acidity and peroxide index, and sensory parameters, such as the fruity median and the median defect. 

\begin{table}[hbt] 
\centering
\caption{Number of olive oils samples in each quality class. EVOO: extra virgin olive oil, VOO: virgin olive oil, LOO: lampante olive oil.\label{tab:oils}}
\begin{tabular}{lc} \toprule
\textbf{Quality}	& \textbf{Number of samples}  \\
\midrule
         EVOO & 12 \\
         VOO & 8 \\
         LOO & 7 \\
\bottomrule
\end{tabular}
\end{table}

All oils were stored in the dark and at 20 $^{\circ}$C during the entire time of the measurements. For the data acquisition, the samples were placed into commercial transparent 0.4 ml glass vials, taking care that no headspace was present to reduce oxidation \cite{iqdiam2020influence}.
For a few selected oils, two samples were prepared from the same olive oil bottle to check the variability of the samples and no difference was observed.

All the measurements in this work were done on undiluted samples. It is well known, that fluorescence in olive oil is subjected to the inner effect \cite{skoog2017principles}, which includes both the attenuation of the excitation light due to the strong absorption from the sample and the re-absorption of the fluorescence light from the sample itself, due to the overlap of the excitation and emission spectra. However, for the technology described in this work, this effect does not pose a problem. In facts, the fluorescence is intense enough that the strong absorption does not influence the signal-to-noise ratio, and possible sample-dependent effects are learned and compensated by machine learning models.

\subsection{Miniaturized Low-Cost Fluorescence Sensor}

\begin{figure}[hbt]
    \centering
    \includegraphics[width=8cm]{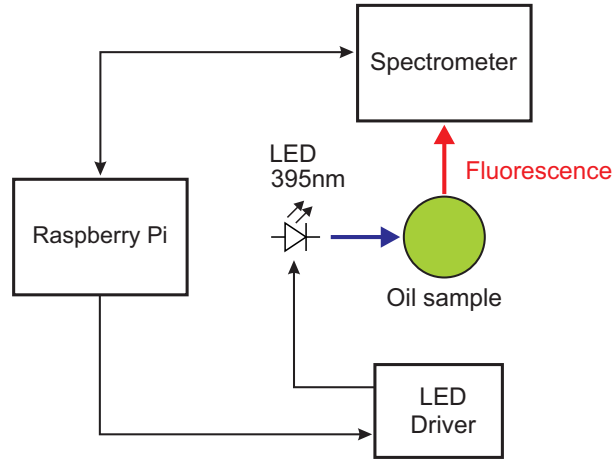}
    \caption{Schematics of the minimalist fluorescence sensor. Blue: excitation light, red: fluorescence light.}
    \label{fig:schematics}
\end{figure}

The design of the sensor was conceived to have as few elements as possible, to minimize the complexity and the costs. For the first time, the sensor itself does not contain any optical component, neither optical filters, as it is typical in fluorescence spectroscopy \cite{lakowicz2013principles}, nor lenses. The schema of the minimalist sensor is shown schematically in Figure \ref{fig:schematics}.

\begin{figure}[hbt]
    \centering
    \includegraphics[width=11cm]{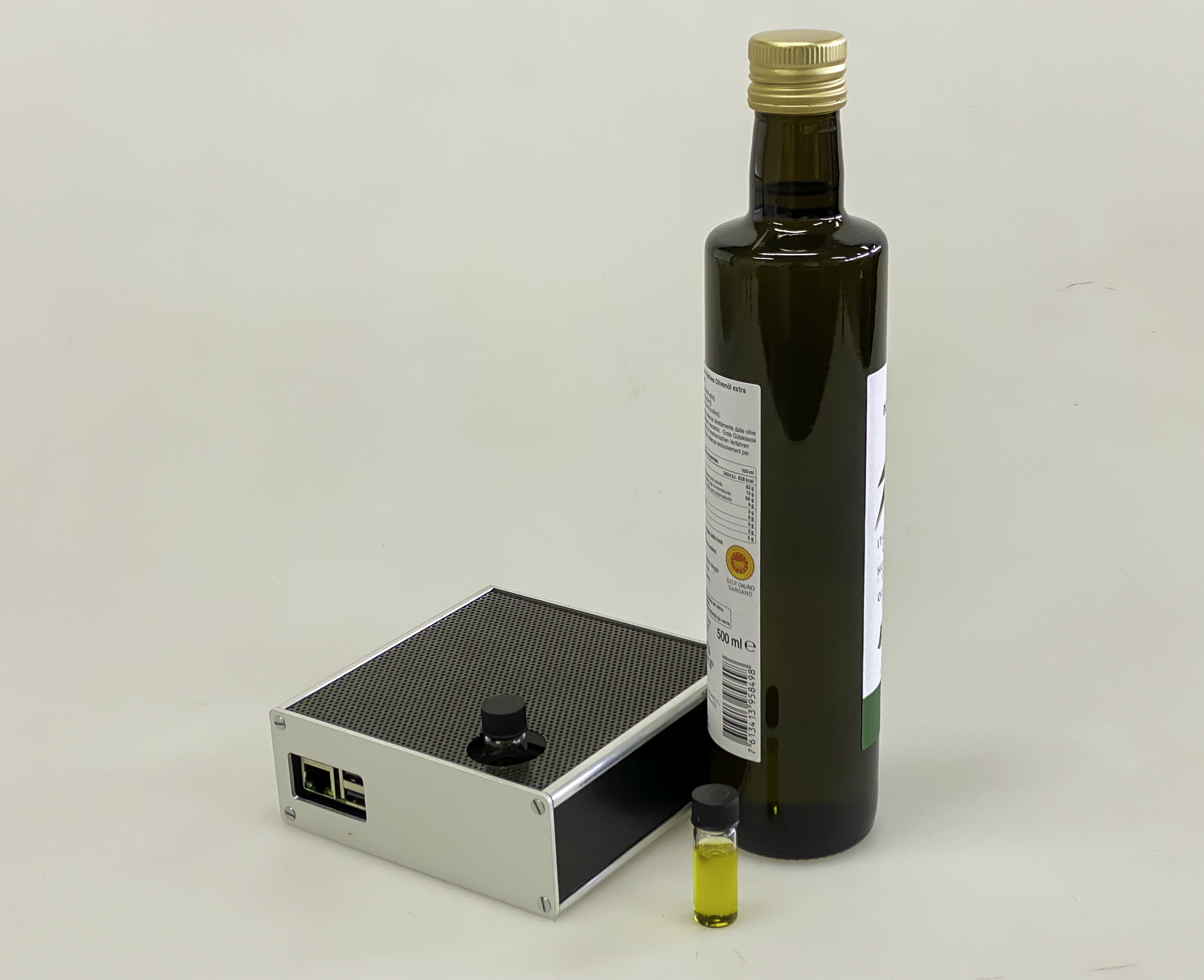}
    \caption{Photo of the minimalist fluorescence sensor with olive oil samples in the glass vials and a bottle of olive oil.}
    \label{fig:photo}
\end{figure}

The excitation light is provided by a UV LED with emission at 395 nm (Kingbright Electronic Co, New Taipei City, Taiwan), driven by a current driver (MIC4801, Micrel Inc., San Jose, CA, USA) which allows regulating the current and, therefore, the illumination intensity. This excitation wavelength is advantageous because it is close to an absorption maximum in the absorption band of the different pigments present in olive oil, mainly chlorophylls and carotenoids \cite{ferreiro2017authentication,torreblanca2019laser,borello2019determination}. The fluorescence is collected by a miniature spectrometer (STS-Vis, Ocean Optics, Dunedin, FL, USA) with a 1024-element CCD array which acquires the entire spectrum in one single measurement. The resolution of the spectrometer is 16 nm. The spectrometer is placed at 90$^\circ$ with respect to the LED to avoid the LED light transmitted by the sample to reach the spectrometer. Both the LED driver and the spectrometer are controlled by a Raspberry Pi.
The optomechanics of the sensor is designed to minimize the amount of stray light from the excitation LED that is collected by the spectrometer.
The current for the LED was chosen so to have a good signal-to-noise ratio for a single spectrum with short integration time avoiding, however, heating the sample with the LED light. The sensor has a recess where standard 0.4 ml clear glass vials can be inserted. The sensor has a very small footprint of 12.5 cm x 12.5 cm x 5 cm and is shown in Figure \ref{fig:photo}. 

\subsection{Dataset Preparation and Description}

All the measurement were taken under ambient conditions in a single day to avoid different aging of the olive oils to influence the results. The description of the samples is reported in Section \ref{sec:OliveOilSamples}. For each olive oil sample, 20 measurements were performed. A total of 27 samples x 20 measurements produced 540 spectra. The dataset, therefore, consists of 540 arrays, each having 1024 values (the number of pixels), whose elements are the measured intensities at the different pixel position after background subtraction, normalized to have an average of zero and a standard deviation of one. This normalization is a very common one with neural networks, as it makes the input data small enough to avoid numerical problems during the training phase \cite{Michelucci2017}.
The dataset contains a different row for each acquisition repetition of each oil, with the spectrum points as features, and the corresponding label for the quality classification.

\subsection{Machine Learning classifiers}
\label{sec:Machinelearningmethods}

The quality of the data acquired with the sensor and the feasibility of using them for quality control were tested by applying different machine learning methods. The goal was to classify the oils into three categories EVOO, VOO, and LOO. The performance of the sensor as a tool for quality control can be defined as its ability to generate data which allow a classification with an accuracy as close to 100$\%$ as possible.
The following eight machine learning algorithms were tested: support vector machines (SVM), na\"ive Bayes (NB), multinomial logistic regression (MLR), PCA combined with LDA, decision tree (DT), ANN, RF, and k-NN. The implementation parameters and the references describing the methods are listed in Table \ref{tab:methods}. The methods have been implemented using the Python library scikit-learn \cite{scikit-learn}. A detailed description of each algorithm goes beyond the scope of this paper, and the interested reader is referred to the listed references. The details of the ANN implementation are described in Subsection \ref{sec:ANNImplementation}.

\begin{table}[hbt] 
\centering
\caption{List of the machine learning methods used, with implementation parameters and references to the 
methods description: support vector machine (SVM), na\"ive Bayes (NB), multi-nomial logistic regression (MLT), principal compnent analysis (PCA) and linear discriminant analysis (LDA), decitions tree (DT), random forest (RF), and k-nearest neighbour (k-NN).\label{tab:methods}}
\begin{tabular}{lp{7cm}l} \toprule
\textbf{Algorithm}	& \textbf{Implementation Details} & \textbf{References}  \\
\midrule
SVM & Regularization parameter $C=1.0$, kernel = radial basis function& \cite{hearst1998support,huang2014introduction,Platt99probabilisticoutputs} \\ 
NB & None & \cite{huang2014introduction,rish2001empirical,islam2007investigating,berrar2018bayes} \\
MLR & regularization penalty = l2, solver algorithm = Newton conjugate gradient & \cite{huang2014introduction,bewick2005statistics,lemeshow1982review,van1988logistic} \\ PCA+LDA	 &Number of components used with LDA: 2, 3, 4, 5, 10, 15, 20, 25 and 30 & \cite{huang2014introduction,martinez2001pca,tominaga1999comparative} \\ 
DT & Split quality criterion used = Gini impurity & \cite{1522531,huang2014introduction,safavian1991survey,song2015decision,swain1977decision} \\
RF & Number of trees = 100, split quality criterion used =  Gini impurity & \cite{1522531,huang2014introduction,ho1995random,ho1998random,RF1} \\ 
k-NN & Numbers of neighbours $k = 3$ &\cite{huang2014introduction,hodges1950discriminatory,kNN1} \\ 

\bottomrule
\end{tabular}
\end{table}

\subsection{Artificial Neural Network-based Classifiers}
\label{sec:ANNImplementation}

For the oil classification, a feed-forward neural network architecture was used. To find the best parameters of the neural network's model (NNM), namely the number of layers, the number of neurons in each layer and the number of epochs, a hyperparameter optimization was performed with a grid-search approach \cite{Michelucci2017}. The number of layers was varied from one to three, the number of neurons in each layer from two to 32 and the number of epochs tested were 350, 600, and 1000. The activation function of the hidden layer's neurons is the rectified linear unit (ReLU) \cite{Michelucci2017}
\begin{equation}
    \textrm{ReLU(x)} \equiv \max \{0,x\}
\end{equation}
while for the output layer the softmax \cite{Michelucci2017} function was used. 
The loss function used is the cross-entropy \cite{Michelucci2017}
\begin{equation}
    L=- \sum_i \sum_{j=1}^3 \left( 
        y_i \log \hat y_i + (1-y_i) \log (1-\hat y_i)
    \right)
\end{equation}
where the sum over $i$ is performed over all the observations on the mini-batch extracted from the training dataset used for the weight update, $y_i$ is the expected class and $\hat y_i$ is the predicted probability of the observation of being of the $j^{th}$ class. $j=1,2,3$ indicates the three expected classes: EVOO, VOO, and LOO.

The NNM was trained using the optimizer Adaptive Moment Estimation (Adam) \cite{kingma2017adam} with a mini-batch size of 32.
The implementation was performed using the TensorFlow~$^{TM}$ Python library.
As will be discussed in the results section, the NNM that gave the best performance was the one with three layers, 32 neurons in each layer, and was trained for 1000 epochs.
 
To measure the performance of the models, the accuracy calculated as the number of correctly classified oils divided by the total number of oils was used.
All the models were trained with backpropagation.

\subsection{External Validation of Models}
\label{sec:validation}

To assess the performance of the machine learning models, these need to be applied to  data not used during training and the resulting prediction tested against the expected results. For this purpose, the dataset was split into 80\%, used as training dataset, and 20\% used for validation \cite{Michelucci2017, huang2014introduction}. All the results reported in this work were obtained on the validation portion of the dataset. The accuracy is defined as the percentage of the olive oils of the validation dataset which are correctly classified. Since variation in the accuracy may arise from the specific split which was performed, the split and train process needs to be repeated several times \cite{michelucci2021estimating}. In this work, the split and train process was repeated 100 times for all algorithms. Then, for all the methods the the average and standard deviation of the accuracy over 100 splits were calculated. These are the result described in Section \ref{sec:resultsclass}. 

\section{Results and Discussion}

In this section firstly the results of the measurements are presented. Finally, the results of the classification using the different techniques are reported.

\subsection{Spectral Response of the Olive Oils}

\begin{figure}[t!]
    \centering
    \includegraphics[width=10cm]{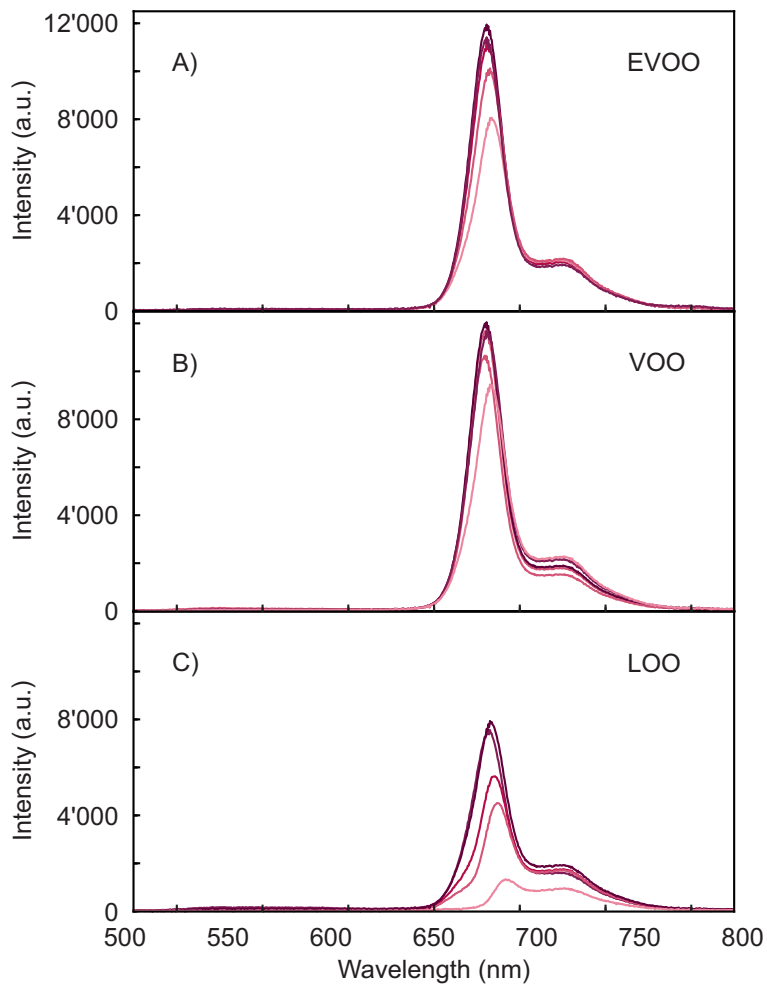}
    \caption{Fluorescence emission spectra of selected olive oils. Panel A): five EVOOs, panel B) five VOOs; Panel C) five LOOs. Each curve shows a single spectrum without averaging or smoothing.}
    \label{fig:Spectra}
\end{figure}

The raw fluorescence spectra of selected EVOOs, VOOs, and LOOs are shown in Figure \ref{fig:Spectra}. In all the figures the curves are just one single spectrum with the background subtracted, without averaging or smoothing. The integration time was 1 second.

Panel A) of Figure \ref{fig:Spectra} shows the fluorescence spectra of EVOOs. For clarity, the spectra of only five of the 12 oils are plotted.
The spectra are characterized by a strong signal in the region between 650 nm and 750 nm, with an intense peak at ca. 678 nm and a weaker broader one at ca. 720 nm, typical of chlorophyll and pheophytins \cite{hernandez2017fast,mishra2018monitoring,baltazar2020development,galeano2003simultaneous}. The stronger peak has not always the same spectra position and intensity, while the broader one is weakly varying between the samples. These observations are consistent with those previously reported and are attributed to the inner filter effect \cite{torreblanca2019laser}.

Noticeably, the spectra below 650 nm do not show any significant fluorescence intensity. This spectral region is usually attributed to underlying chemical constituents, such as vitamin E, hydrolysis and oxidation products \cite{kongbonga2011characterization,baltazar2020development}. The lack of significant fluorescence signal in this region is due to the choice of the excitation wavelength. These compounds absorb in the UV, well below the excitation wavelength peaked at 395 nm used here. Depending on the sensor purpose, the inclusion of an additional UV LED to acquire also the UV fluorescence contributions th the spectrum could increase the performance by providing additional specific information. For the problem studied in this work, the strong fluorescence contribution between 650 nm and 750 nm proved to be enough to achieve 100$\%$ classification.

For comparison the fluorescence spectra of VOOs and LOOs are shown in Figure \ref{fig:Spectra} panels B) and C). The VOOs show emission spectra which are similar to the EVOOs, with a stronger variability particularly in the intensity of the broader shoulder at 720 nm. The variability between the spectra increases further in the LOOs.
The fluorescence from EVOO and VOO samples is generally stronger than LOO ones, consistently with what previously reported observations for LOOs obtained with synchronous fluorescence spectroscopy \cite{poulli2005classification}.

\subsection{Classification with Machine Learning Methods}
\label{sec:resultsclass}

The results of the classification with all the machine learning methods are summarized in Table \ref{tab:mlalgos}. The results are given as the average of the accuracy $\overline a$ over 100 different splits and the standard deviation of the accuracy, as described in Section \ref{sec:validation}.

\begin{table}[hbt] 
\centering
\caption{Summary of results of the classification given by the average of the accuracy $\overline a$ and its standard deviation $\sigma$; machine learning methods: support vector machine (SVM), na\"ive Bayes (NB), multi-nomial logistic regression (MLT), principal compnent analysis (PCA) and linear discriminant analysis (LDA), decitions tree (DT), random forest (RF), and k-nearest neighbour (k-NN).
\label{tab:mlalgos}}
\begin{tabular}{rcc} \toprule
\textbf{Algorithm}	& Average accuracy & Standard deviation \\
	& $\overline a$ & $\sigma$ \\
\midrule
SVM	& 0.51	& 0.07\\
NB	& 0.64	& 0.05\\
MLR & 0.88 & 0.03\\
PCA + LDA & 0.93 & 0.02\\ 
(10 PCA Components) & & \\
DT & 0.99 & 0.01\\
ANN & 0.99 & 0.04 \\
PCA + LDA & 0.999 & 0.006\\ 
(30 PCA Components) & & \\
RF & 1.0 & 0.0\\ 
k-NN & 1.0 & 0.0\\ 
\bottomrule
\end{tabular}
\end{table}

As it can be seen from Table \ref{tab:mlalgos} several methods allow reaching an average accuracy above 99$\%$ without any pre-processing, namely the DT, ANN, PCA combined with LDA, RF, and k-NN. There results are better than previously reported for the classification between VOO and LOO with Hierarchical Cluster Analysis (HCA) on EEMs and similar to what obtained with PCA \cite{poulli2005classification}. Unsurprisingly, the results obtained with SVM are poorer than those obtained with the other methods as typically with those algorithms pre-processing is a key part of the analysis. In fact, in previous work, SVM was applied after pre-processing the data, for example with PCA, to obtain a good accuracy \cite{el2020rapid}. PCA with LDA was studied using increasing number of PCA components: 2, 3, 4, 5, 10, 15, 20, 25 and 30. Already by using only 10 PCA components, LDA was able to reach an accuracy over 90\%. With 30 the accuracy reached was over 99\%. It is important to note that each spectrum (input to the PCA) consists of 1024 values (the pixels of the CCD of the spectrometer), thus using 30 PCA components is equivalent to using only 2.9\% of the amount of features in the original spectra.

\begin{figure}[hbt]
    \centering
    \includegraphics[width=13cm]{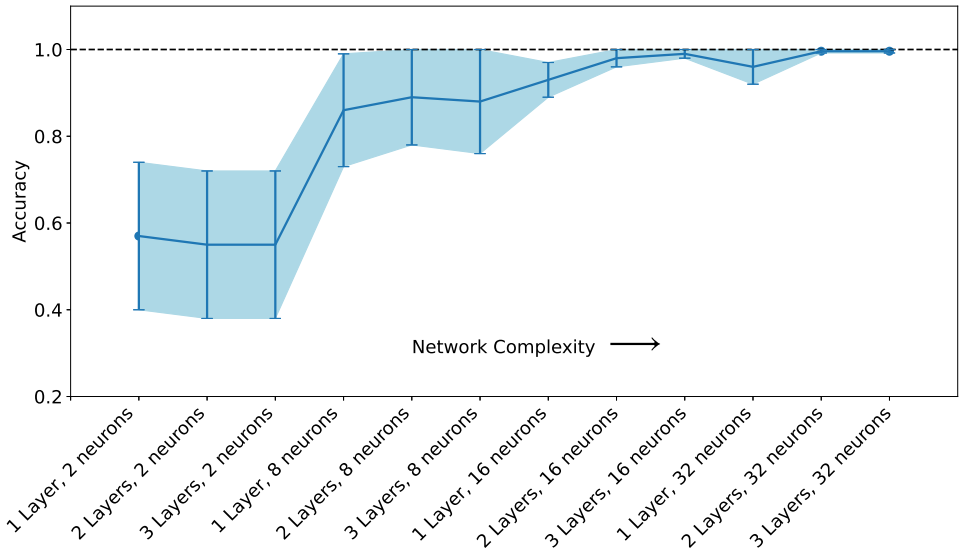}
    \caption{Evolution of the average of the accuracy and its standard deviation with with ANN's complexity. For each architecture the points indicate the average of the accuracy of 100 split and train runs, and the error lines indicate the standard deviation.}
    \label{fig:hpt}
\end{figure}

To find the optimal architecture for the ANN, hyperparameter tuning was performed as described in Section \ref{sec:ANNImplementation}. The evolution of the average of accuracy and its standard deviation with increasing ANN's complexity is shown in Figure \ref{fig:hpt}. The vertical bar indicates the standard deviation calculated from the 100 different splits. Only the results obtained with 1000 epochs are shown. The effect of increasing the number of epochs from 350 to 1000 was to improve the accuracy and reduce the standard deviation of the average of the accuracy, as expected. Above 1000 epochs the performance increase is smaller than what obtained by changing the number of layers. Since for moderately complex networks the accuracy was 100$\%$ with 1000 epochs, the training was not performed for a larger number of epochs.

For very simple architectures, with only two neurons, the accuracy is below 60$\%$. The use of eight neurons already improves the accuracy to above 80$\%$. When using 32 neurons the accuracy is always above 90$\%$, and increases to above 99$\%$ when using two layers. The increase from two to three layer does not affect the results significantly, being the accuracy already practically 100$\%$. This means that the ANN can always correctly identify the three classes of olive oil quality (EVOO, VOO, and LOO).

The goal of this work is to demonstrate that the fluorescence sensor is able to generate data that can be used without any pre-processing or manual feature engineering to make the classification process as easy and automatic as possible. As it can be seen from Table \ref{tab:mlalgos} this is the case. These results indicate without any doubt that the data acquired with this very simple and low-cost spectrometer contain sufficient information to allow the correct discrimination between the three quality classes with almost perfect accuracy.

\section{Conclusions}

The current work presents a new type of compact and low-cost fluorescence sensor which allows high-quality data acquisition that can be reliably used for data-processing or inference for classification purposes. The sensor is extremely simply conceived to minimize its size and costs and allow portability.
The results demonstrate the use of a minimalist optical sensor based on fluorescence spectroscopy associated with machine-learning methods that can reliably distinguish between different qualities of olive oil: EVOO, VOO, and LOO. This new low-cost sensor has the advantage of being a portable, easy-to-use and low-cost device, which works with undiluted samples, without any handling of the olive oils, like dilution, and without any pre-processing of the data, thus simplifying the analysis to the maximum degree possible. Problems like strong absorption and inner filter effect do not affect the performance because they are learnt and compensated by the machine learning methods.
Among the methods, the use of ANN is particularly important because it does not require pre-processing of the data and allows the use of flexible explainability techniques to better optimize and understand the classification process.

The problem investigated here is just one example of the many possible applications. The sensor can be used to solve other classification and regression problems. The details of the machine learning models are expected to be specific of the problem to be addressed.

\noindent
{{\bf Authors Contribtions: }Conceptualization and methodology F.V., U.M. and M.A.D.; software U.M and M.S.; data curation M.S.; hardware M.B and I.H.; writing - original draft preparation F.V.; writing - review and editing F.V., U.M. and M.S. All authors have read and agreed to the published version of the manuscript.}

\noindent
{{\bf Funding: }This research was partially funded by Innosuisse - Swiss Innovation Agency, Grant No.: 36761.1 INNO-LS and the Virtuous project, funded by the European Union’s Horizon 2020 research and innovation programme under the Maria Sklodowska-Curie—RISE Grant Agreement No 872181.}

\bibliographystyle{unsrt}  
\bibliography{main}  

\end{document}